# Performance Analysis of 3-Dimensional Turbo Codes

Eirik Rosnes, *Senior Member, IEEE,* and Alexandre Graell i Amat, *Senior Member, IEEE*

*Abstract*—In this work, we consider the minimum distance properties and convergence thresholds of 3-dimensional turbo codes (3D-TCs), recently introduced by Berrou *et al.*. Here, we consider binary 3D-TCs while the original work of Berrou *et al.* considered double-binary codes. In the first part of the paper, the minimum distance properties are analyzed from an ensemble perspective, both in the finite-length regime and in the asymptotic case of large block lengths. In particular, we analyze the asymptotic weight distribution of 3D-TCs and show numerically that their typical minimum distance $d_{\min}$ may, depending on the specific parameters, asymptotically grow linearly with the block length, i.e., the 3D-TC ensemble is asymptotically good for some parameters. In the second part of the paper, we derive some useful upper bounds on the $d_{\min}$ when using quadratic permutation polynomial (QPP) interleavers with a quadratic inverse. Furthermore, we give examples of interleaver lengths where an upper bound appears to be tight. The best codes (in terms of estimated $d_{\min}$) obtained by randomly searching for good pairs of QPPs for use in the 3D-TC are compared to a probabilistic lower bound on the $d_{\min}$ when selecting codes from the 3D-TC ensemble uniformly at random. This comparison shows that the use of designed QPP interleavers can improve the $d_{\min}$ significantly. For instance, we have found a $(6144, 2040)$ 3D-TC with an estimated $d_{\min}$ of $147$, while the probabilistic lower bound is $69$. Higher rates are obtained by puncturing nonsystematic bits, and optimized periodic puncturing patterns for rates $1/2$, $2/3$, and $4/5$ are found by computer search. Finally, we give iterative decoding thresholds, computed from an extrinsic information transfer chart analysis, and present simulation results on the additive white Gaussian noise channel to compare the error rate performance to that of conventional turbo codes.

## I. Introduction

Turbo codes have gained considerable attention since their introduction by Berrou *et al.* in 1993 [1] due to their near-capacity performance and low decoding complexity. The conventional turbo code is a parallel concatenation of two identical recursive systematic convolutional encoders separated by a pseudo-random interleaver. Most conventional turbo codes using 8-state constituent encoders suffer from a flattening around a frame error rate (FER) of $10^{-5}$ due to a poor minimum distance $d_{\min}$. To improve the performance in the error floor region, one could either design a better interleaver, use more powerful constituent encoders, or increase the dimension, i.e., the number of constituent encoders. The latter alternative was recently pursued in [2, 3], where a powerful coding scheme nicknamed 3-dimensional turbo code (3D-TC) and inspired by the proposals in [4, 5], was introduced. The coding scheme consists of a conventional turbo encoder and a *patch*. In more detail, a fraction $\lambda$ of the parity bits from the turbo encoder are post-encoded by a third rate-1 encoder. The value of $\lambda$ can be used to trade-off performance in the waterfall region with performance in the error floor region. As shown in [2, 3], this coding scheme is able to provide very low error rates for a wide range of block lengths and code rates at the expense of a small increase in decoding complexity with respect to conventional turbo codes.

It is known that conventional turbo codes and single serially concatenated convolutional codes are asymptotically bad, in the sense that their typical $d_{\min}$ asymptotically does not grow linearly with the block length [6, 7]. As an alternative, multiple serially concatenated codes, such as repeat multiple-accumulate (RMA) codes, can be used, since they yield a better $d_{\min}$. In [8], it was shown that there exists a sequence of RMA codes with minimum distance converging in the limit of infinitely many accumulators to the Gilbert-Varshamov bound (GVB). The stronger result that the typical $d_{\min}$ converges to the GVB was recently proved in [9]. Also, in [10], it was conjectured by Pfister that the $d_{\min}$ of RMA codes asymptotically grows linearly with the block length, and that the growth rate is given by the threshold where the asymptotic spectral shape function [11] becomes positive. Kliewer *et al.* and Ravazzi and Fagnani showed independently in [9, 12] that RMA code ensembles with two or more accumulators are indeed asymptotically good (their typical $d_{\min}$ asymptotically grows linearly with the block length). A formal proof was given in [9], and a method for the calculation of a lower bound on the asymptotic growth rate coefficient was given in [12]. The analysis was later extended in [13, 14] to low-rate hybrid concatenated codes, i.e., mixed parallel and serial structures combining the features of the two concatenations.

In the first part of this paper, following these works, we analyze the minimum distance properties of 3D-TCs. We perform an asymptotic $d_{\min}$ analysis of 3D-TCs by using a numerical procedure to estimate their asymptotic spectral shape function. The numerical procedure is based on the approach in [15] to compute asymptotic input-output weight distributions of

The material in this paper was presented in part at the 5th International Symposium on Turbo Codes & Related Topics, Lausanne, Switzerland, September 2008, and at the IEEE Information Theory Workshop, Cairo, Egypt, January 2010. The work of E. Rosnes was supported by the Research Council of Norway (NFR) under Grants 174982 and 183316. A. Graell i Amat was supported by a Marie Curie Intra-European Fellowship within the 6th European Community Framework Programme.

E. Rosnes is with the Selmer Center, Department of Informatics, University of Bergen, N-5020 Bergen, Norway (e-mail: eirik@ii.uib.no).

A. Graell i Amat is with the Department of Signals and Systems, Chalmers University of Technology, Gothenburg, Sweden (e-mail: alexandre.graell@chalmers.se).

convolutional encoders. It is shown numerically that for certain parameters, the $d_{\min}$ of 3D-TCs asymptotically grows linearly with the block length. We also perform a finite-length analysis of the $d_{\min}$ of 3D-TCs using a probabilistic lower bound, outlined in [8], on the $d_{\min}$.

Interleavers for conventional turbo codes have been extensively investigated. The dithered relative prime (DRP) interleavers [16, 17] and the almost regular permutation (ARP) interleavers [18] are considered among the best ones, since they provide relatively high $d_{\min}$. Recently, Sun and Takeshita [19] suggested the use of permutation polynomial (PP) based interleavers over integer rings. In particular, quadratic polynomials were emphasized. In contrast to DRP and ARP interleavers, PP interleavers are fully algebraic, allowing a theoretical analysis of their performance. In [20], the $d_{\min}$ of conventional binary turbo codes with quadratic permutation polynomial (QPP) interleavers was considered in detail, and large tables of optimum (in terms of $d_{\min}$ and its corresponding multiplicity) QPPs for conventional turbo codes with 8-state and 16-state constituent encoders were presented. In the most recent work [21], Takeshita considered the use of higher degree PPs with great success.

A suitable property for designed interleavers is the *contention-free* property, i.e., to avoid memory contentions in parallelized decoding [18, 22, 23]. While ARP and some modified DRP interleavers [22] are contention-free, they are not maximum contention-free, i.e., every factor of the interleaver length is not a possible degree of parallel processing of the decoder. On the other hand, in [24] it was shown that all PPs generate maximum contention-free interleavers. Thus, these interleavers are very interesting from an implementation point of view. Furthermore, QPP interleavers are almost as good as DRP interleavers for a large number of short-to-medium block lengths in terms of decoding convergence and performance in the error floor region [19–21, 24].

In the second part of this work, we analyze minimum distance properties of 3D-TCs with dedicated QPP interleavers. In particular, we present several upper bounds on the $d_{\min}$ of binary 3D-TCs when using QPP interleavers with a quadratic inverse that do not depend on the permutation and the encoder in the patch, as long as the encoder maps the all-zero sequence to the all-zero sequence. Furthermore, we present some results from a random search for good pairs of QPPs for use in the binary 3D-TC. It is shown that the use of designed QPPs yields a very high $d_{\min}$, improving significantly compared to the probabilistic lower bound on the $d_{\min}$.

The remainder of this paper is organized as follows. The encoder structure and design guidelines for 3D-TCs are described in Section II. Section III describes a probabilistic lower bound on the $d_{\min}$ of a code ensemble and its application to 3D-TCs. The asymptotic spectral shape function is introduced in Section IV along with a numerical procedure to estimate it. Furthermore, we show numerically that the 3D-TC ensemble is asymptotically good for certain parameters. The minimum distance properties of 3D-TCs with dedicated QPP interleavers are addressed in Sections V and VI. In particular, Section V describes QPPs and some of their properties, and in Section VI, an upper bound on the $d_{\min}$ with QPP interleavers with a quadratic inverse for a conventional turbo code is presented, along with upper bounds on the $d_{\min}$ of 3D-TCs with QPP interleavers with a quadratic inverse and with any patch. Convergence properties are studied in Section VII, where an extrinsic information transfer (EXIT) chart analysis is performed. In Section VIII, the results of a random search for good pairs of QPPs for use in the 3D-TC are presented and compared to the finite-length results from Section III. Also, optimized puncturing patterns for rates $1/2$, $2/3$, and $4/5$ are established by computer search. Finally, simulation results on the additive white Gaussian noise (AWGN) channel are presented to compare the error rate performance to that of conventional turbo codes. Conclusions and a discussion of future work are given in Section IX.

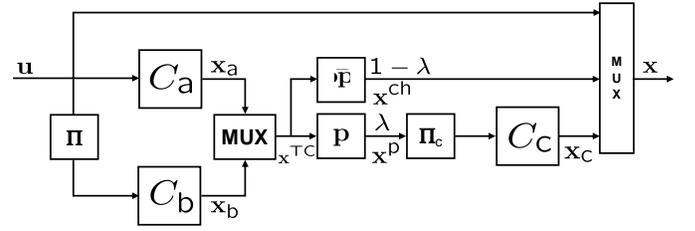

Fig. 1. 3D turbo encoder. A fraction $\lambda$ of the parity bits from both constituent encoders $C_a$ and $C_b$ are grouped by a parallel/serial multiplexer, permuted by interleaver $\Pi_c$, and encoded by the rate-1 post-encoder $C_c$.

## II. CODING SCHEME

A block diagram of the 3D-TC is depicted in Fig. 1. The information data sequence $\mathbf{u}$ of length $K$ bits is encoded by a binary conventional turbo encoder. By a conventional turbo encoder we mean the parallel concatenation of two identical rate-1 recursive convolutional encoders, denoted by $C_a$ and $C_b$, respectively. Here $C_a$ and $C_b$ are 8-state recursive convolutional encoders with generator polynomial $g(D) = (1 + D + D^3)/(1 + D^2 + D^3)$, i.e., the 8-state constituent encoder specified in the 3GPP/UMTS standard [25]. The code sequences of $C_a$ and $C_b$ are denoted by $\mathbf{x}_a$ and $\mathbf{x}_b$, respectively. We also denote by $\mathbf{x}^{\text{TC}}$ the codeword obtained by alternating bits from $\mathbf{x}_a$ and $\mathbf{x}_b$. A fraction $\lambda$ ($0 \leq \lambda \leq 1$), called the *permeability rate*, of the parity bits from $\mathbf{x}^{\text{TC}}$ are permuted by interleaver $\Pi_c$ (of length $N_c = 2\lambda K$), and encoded by an encoder of unity rate $C_c$, called the *patch* or the *post-encoder* [3]. This can be properly represented by a puncturing pattern $\mathbf{p}$ applied to $\mathbf{x}^{\text{TC}}$ (see Fig. 1) of period $N_p$ containing $\lambda N_p$ ones (where a one means that the bit is not punctured). The fraction $1 - \lambda$ of parity bits which are not encoded by $C_c$ is sent directly to the channel. Equivalently, this can be represented by a puncturing pattern $\bar{\mathbf{p}}$, the complement of $\mathbf{p}$. We denote by $\mathbf{x}_c$ the code sequence of $C_c$. Also, we denote by $\mathbf{x}_a^{\text{ch}}$ and $\mathbf{x}_b^{\text{ch}}$ the *sub-codewords* of $\mathbf{x}_a$ and $\mathbf{x}_b$, respectively, sent directly to the channel, and by $\mathbf{x}^{\text{ch}}$ the codeword obtained by alternating bits from $\mathbf{x}_a^{\text{ch}}$ and $\mathbf{x}_b^{\text{ch}}$. Likewise, we denote by $\mathbf{x}_a^{\text{p}}$ and $\mathbf{x}_b^{\text{p}}$ the *sub-codewords* of $\mathbf{x}_a$ and $\mathbf{x}_b$, respectively, encoded by $C_c$, and by $\mathbf{x}^{\text{p}}$ the codeword obtained by alternating bits from $\mathbf{x}_a^{\text{p}}$ and $\mathbf{x}_b^{\text{p}}$. Finally, the information sequence and the code sequences $\mathbf{x}^{\text{ch}}$ and $\mathbf{x}_c$ are multiplexed to form the code sequence **x**, of length $N$ bits, transmitted to the channel. Note that the overall nominal code rate of the 3D-TC is $R = K/N = 1/3$, the same as for the conventional turbo code without the patch. Higher code rates can be obtained either by puncturing $\mathbf{x}^{\text{ch}}$ or by puncturing the output of the patch, $\mathbf{x}_c$. In this paper, we consider the following puncturing strategy. First, puncture $\mathbf{x}^{\text{ch}}$. Then, if further puncturing is required, puncture $\mathbf{x}_c$.

In [3], regular puncturing patterns of period $2/\lambda$ were considered for **p**. For instance, if $\lambda = 1/4$, every fourth bit from each of the encoders of the outer turbo code are encoded by encoder $C_c$. The remaining bits are sent directly to the channel, and it follows that $\mathbf{p} = [11000000]$ and $\bar{\mathbf{p}} = [00111111]$. In this paper, we consider both regular and random patterns for **p**.

The 3D-TC can be decoded using the turbo principle. The decoder consists of three soft-input soft-output decoders $C_a^{-1}$, $C_b^{-1}$, and $C_c^{-1}$ corresponding to the three constituent encoders $C_a$, $C_b$, and $C_c$, respectively. A decoding iteration consists of a single activation of $C_c^{-1}$, $C_a^{-1}$, and $C_b^{-1}$, in this order. This process continues iteratively until the maximum number of iterations is reached or an early stopping rule criterion is fulfilled.

*A. Design Guidelines*

In [3], some guidelines for choosing the permeability rate $\lambda$ were given. In general, choosing a large value for $\lambda$ will increase the minimum distance. However, the performance in the waterfall region will degrade with increasing values of $\lambda$. Thus, there is a trade-off between performance in the waterfall and error floor regions.

According to [3], the choice of the post-encoder is crucial for the code performance. In general, the post-decoder must be simple and be able to handle soft-input and produce soft-output information. Furthermore, the post-decoder must not exhibit too much *error amplification* (see [3] for details), since this will result in a high loss in convergence. In this paper, we consider the encoder with generator polynomial $g(D) = 1/(1 + D^2)$ for $C_c$ [3].

## III. WEIGHT ENUMERATORS AND FINITE-LENGTH MINIMUM DISTANCE ANALYSIS

In this section, we analyze the minimum distance properties of 3D-TCs. In particular, we consider the ensemble of codes in the form of Fig. 1 obtained by considering all possible permutations for $\Pi$ and $\Pi_c$ through the *uniform interleaver* approach [26].

Let $A_{w,h}^C$ denote the input-output weight enumerator (IOWE) of a code $C$, i.e., the number of code sequences of weight $h$ corresponding to input sequences of weight $w$. Also, let $A_h^C = \sum_w A_{w,h}^C$ be the weight enumerator (WE) of the code $C$, i.e., the number of code sequences of weight $h$. Denote by $q_a$ and $q_b$ the weights of code sequences $\mathbf{x}_a$ and $\mathbf{x}_b$, respectively, and by $q$ its sum, $q = q_a + q_b$ (i.e., the weight of $\mathbf{x}^{\text{TC}}$). We also denote by $n_a$, $n_b$, and $n$ the weights of code sequences $\mathbf{x}_a^p$, $\mathbf{x}_b^p$, and $\mathbf{x}^p$, respectively ($n = n_a + n_b$). Likewise, we denote by $m_a$, $m_b$, and $m$ the weights of code sequences $\mathbf{x}_a^{\text{ch}}$, $\mathbf{x}_b^{\text{ch}}$, and $\mathbf{x}^{\text{ch}}$, respectively ($m = m_a + m_b$). Note that $q = n + m$.

*A. IOWE of 3D-TCs With Random Puncturing Pattern* **p**

We assume a random puncturing pattern for **p**. The puncturing patterns are sampled uniformly at random among all those with $\lfloor \delta N \rfloor$ ones, where $\delta$ is the fraction of bits that survive after puncturing. The average IOWE of the ensemble of punctured codes $\mathcal{C}_{\text{punct}}$, with input weight $w$ and output weight $h'$ is given by [12]

$$\bar{A}_{w,h'}^{\mathcal{C}_{\text{punct}}} = \sum_{h=h'}^{N} A_{w,h}^C \frac{\binom{h}{h'}\binom{N-h}{\lfloor \delta N \rfloor - h'}}{\binom{N}{\lfloor \delta N \rfloor}} \quad (1)$$

where $h$ is the output weight before puncturing.

Using the concept of uniform interleaver [26] and (1) the ensemble-average IOWE of the 3D-TC ensemble, denoted by $\mathcal{C}$, can be computed as

$$\bar{A}_{w,h}^{\mathcal{C}} = \sum_{q,q_a,n} \frac{A_{w,q_a}^{C_a} A_{w,q-q_a}^{C_b}}{\binom{K}{w}} \cdot \frac{\binom{q}{n}\binom{2K-q}{2\lambda K-n}}{\binom{2K}{2\lambda K}} \cdot \frac{A_{n,h-w-q+n}^{C_c}}{\binom{2\lambda K}{n}}. \quad (2)$$

*B. IOWE of 3D-TCs With Regular Puncturing Pattern* **p**

Here, we assume the use of regular (i.e., nonrandom) puncturing patterns for **p**. In this case, the ensemble-average IOWE of the 3D-TC ensemble can be written as

$$\bar{A}_{w,h}^{\mathcal{C}} = \sum_{m,m_a,n,n_a} \frac{A_{w,(m_a,n_a)}^{C_a} A_{w,(m-m_a,n-n_a)}^{C_b}}{\binom{K}{w}} \cdot \frac{A_{n,h-w-m}^{C_c}}{\binom{2\lambda K}{n}} \quad (3)$$

where $A_{w,(m_x,n_x)}^{C_x}$, $x = a, b$, is the number of codewords of constituent encoder $C_x$ with input weight $w$, and output weights $m_x$ and $n_x$ corresponding to the sub-codewords sent directly to the channel and to encoder $C_c$, respectively. We remark that the two enumerators in (2) and (3) are not the same, since the first is averaged over puncturing patterns, while the second is for a given puncturing pattern. Thus, there is a slight abuse of notation.

*C. Finite-Length Minimum Distance Analysis*

The ensemble-average WE $\bar{A}_h^{\mathcal{C}}$ can be used to bound the minimum distance $d_{\min}$ of the code ensemble $\mathcal{C}$ in the finite-length regime. In particular, the probability that a code randomly chosen from the ensemble has minimum distance $d_{\min} < d$ is upper-bounded by [8]

$$\Pr(d_{\min} < d) \leq \sum_{h=1}^{d-1} \bar{A}_h^{\mathcal{C}}. \quad (4)$$

The upper bound in (4) can be used to obtain a probabilistic lower bound on the minimum distance of a code ensemble. For a fixed value of $\epsilon$, where $\epsilon$ is any positive value between 0 and 1, we define the probabilistic lower bound with probability $\epsilon$, denoted by $d_{\min,\text{LB},\epsilon}$, to be the largest integer $d$ such that the right-hand side of (4) is at most $\epsilon$. This guarantees that $\Pr(d_{\min} \geq d) \geq 1 - \epsilon$.

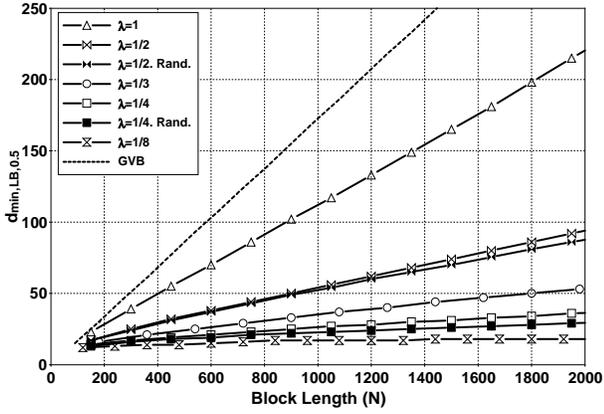

Fig. 2. Probabilistic lower bound on the minimum distance of 3D-TC ensembles, $R = 1/3$, for several values of $\lambda$ using regular and random puncturing patterns, when $\epsilon = 0.5$.

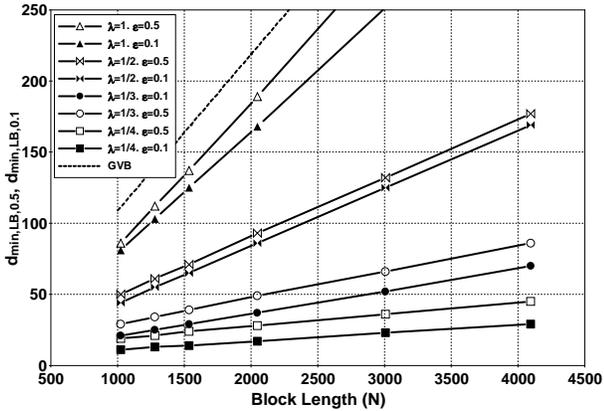

Fig. 3. Probabilistic lower bound on the minimum distance of 3D-TC ensembles, $R = 1/2$, for several values of $\lambda$ using regular puncturing patterns, when $\epsilon = 0.1$ and $0.5$.

In Fig. 2, we plot the probabilistic lower bound from (4), $d_{\min,\text{LB},0.5}$, for 3D-TC ensembles, $R = 1/3$, with regular puncturing patterns $\mathbf{p}$ and for several values of $\lambda$ as a function of the block length $N$. We assumed $\epsilon = 0.5$, which implies that $\Pr(d_{\min} \geq d) \geq 0.5$, i.e., we expect that at least half of the codes in $\mathcal{C}$ have a $d_{\min}$ at least equal to the value predicted by the curves. For comparison purposes, the GVB is also displayed in the figure. For values of $\lambda$ down to $1/4$, the growth rate of the $d_{\min}$ appears to be linear with the block length, while for smaller values of $\lambda$ little growth is observed. The achievable minimum distances are quite high, especially for high values of $\lambda$. For instance, the bound for the 3D-TC ensemble with $\lambda = 1$ is $d_{\min} \approx 215$ for a block length of $N = 1950$ bits. Reducing $\lambda$, significantly reduces the growth rate. For instance, for $\lambda = 1/2$, the probabilistic lower bound gives $d_{\min} \approx 92$ for the same block length. For higher rates, the curves get closer to the GVB. In Fig. 2, we also plot the probabilistic lower bound from (4) for $R = 1/3$ 3D-TC ensembles with $\lambda = 1/2$ and $1/4$ using random puncturing patterns $\mathbf{p}$. Compared to regular puncturing patterns, the growth rate of the $d_{\min}$ is slightly smaller. Thus, regular puncturing patterns seem to be a good choice for $\mathbf{p}$.

In Fig. 3, we plot the probabilistic lower bound from (4) for $R = 1/2$ 3D-TC ensembles using regular puncturing patterns $\mathbf{p}$ for several values of $\lambda$ and the code block lengths used for the QPP interleavers in Section VIII. In the figure, we plot the bound for $\epsilon = 0.5$ and $0.1$, i.e., $50\%$ and $90\%$ of the codes have minimum distance at least equal to the value predicted by the curves, respectively. In all cases, a linear-like growth rate is observed.

## IV. ASYMPTOTIC MINIMUM DISTANCE ANALYSIS

In this section, we analyze the asymptotic behavior of the WE of 3D-TC ensembles to show that their typical $d_{\min}$ grows linearly with the block length for some parameters. To this end, we consider the behavior of the *asymptotic spectral shape function* of the code ensemble, defined as [11]

$$r(\rho) = \limsup_{N \to \infty} \frac{1}{N} \ln \overline{A}^{\mathcal{C}}_{\lfloor \rho N \rfloor} \tag{5}$$

where $\sup(\cdot)$ denotes the supremum of its argument, $\rho = \frac{h}{N}$ is the normalized output weight, and $N$ is the block length.

From (5), we can write $\overline{A}^{\mathcal{C}}_h \sim e^{Nr(\rho)}$ when $N \to \infty$. Therefore, if there exists some abscissa $\rho_0 > 0$ such that $\sup_{\rho \leq \rho^*} r(\rho) < 0 \quad \forall \rho^* < \rho_0$, and $r(\rho) > 0$ for some $\rho > \rho_0$, then it can be shown, with high probability, that the $d_{\min}$ of most codes in the ensemble grows linearly with the block length $N$, with growth rate coefficient of at least $\rho_0$ [9, 12]. On the other hand, if $r(\rho)$ is strictly zero in the range $(0, \rho_0)$, then we cannot conclude directly whether $d_{\min}$ grows linearly with block length or not. In [9], it was shown that the asymptotic spectral shape function of RMA codes exhibits this behavior, i.e., it is zero in the range $(0, \rho_0)$ and positive for some $\rho > \rho_0$. However, by combining the asymptotic spectral shapes with the use of bounding techniques, Ravazzi and Fagnani were able to prove in [9, Theorem 6] that the minimum distance of RMA codes indeed grows linearly with the block length with growth rate coefficient of at least $\rho_0$.

We remark that in the rest of the paper, with a slight abuse of language, we sometimes refer to $\rho_0$ as the exact value of the asymptotic growth rate coefficient. However, we emphasize that, strictly speaking, $\rho_0$ is only a lower bound on the asymptotic growth rate coefficient.

### A. Asymptotic Spectral Shape Function of 3D-TCs

For analysis purposes, we assume a random puncturing pattern $\mathbf{p}$. Let $C$ be an $(N, K)$ code. We define the asymptotic behavior of the IOWE for $C$ as the function

$$a^C(\alpha, \beta) = \limsup_{N \to \infty} \frac{1}{N} \ln A_{\lfloor \alpha K \rfloor, \lfloor \beta N \rfloor} \tag{6}$$

where $\alpha = w/K$ and $\beta = h/N$ are the normalized input weight and the normalized output weight, respectively. Using (2), (6), and Stirling's approximation for binomial coefficients $\binom{n}{k} \sim e^{n\mathbb{H}(k/n)}$ for $n \to \infty$ and $k/n$ constant, where $\mathbb{H}(\cdot)$ is the binary entropy function with natural logarithms, the asymptotic spectral shape function of the (unpunctured) 3D-

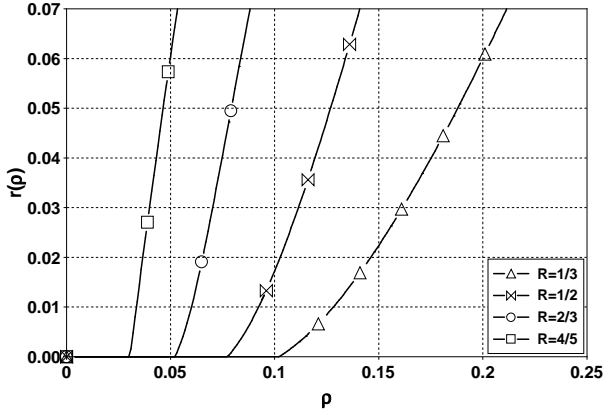

Fig. 4. Asymptotic spectral shape function for 3D-TCs, $\lambda = 1$, and several code rates.

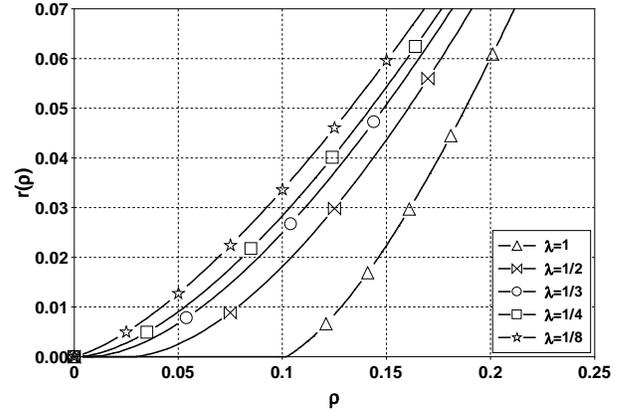

Fig. 5. Asymptotic spectral shape function for $R = 1/3$ 3D-TCs and several values of $\lambda$.

TC ensemble $\mathcal{C}$ can be written as

$$r(\rho) = \frac{1}{3} \sup_{0 \leq \omega, \iota, \iota_a, \mu \leq 1} \left\{ a^{C_a}(\omega, \iota_a) + a^{C_b}(\omega, 2\iota - \iota_a) \right.$$
$$+ 2\iota \mathbb{H}\left(\frac{\lambda \mu}{\iota}\right) + 2(1 - \iota)\mathbb{H}\left(\frac{\lambda(1 - \mu)}{1 - \iota}\right)$$
$$+ 2\lambda a^{C_c}\left(\mu, \frac{3\rho - (\omega + 2(\iota - \lambda \mu))}{2\lambda}\right)$$
$$\left. - \mathbb{H}(\omega) - 2\mathbb{H}(\lambda) - 2\lambda \mathbb{H}(\mu) \right\} \quad (7)$$

where $\omega = w/K$, $\iota_a = q_a/K$, $\iota = q/2K$, and $\mu = n/N_c = n/2\lambda K$. For higher rates, i.e., when puncturing is applied to $\mathbf{x}^{ch}$ or to $\mathbf{x}_c$, a similar expression is obtained with some extra terms. Note that in (7) we did not include the constraints on the relationship of the variables involved in the optimization. The constraints on the involved variables can be derived by looking at the arguments of the involved binary entropy functions in (7). In general, the argument of the binary entropy function should be between 0 and 1.

The asymptotic IOWE of the inner code $C_c$ is equal to the asymptotic IOWE of an accumulate code [15] and can be given in closed form as

$$a^{C_c}(\alpha, \beta) = (1 - \beta)\mathbb{H}\left(\frac{\alpha}{2(1 - \beta)}\right) + \beta \mathbb{H}\left(\frac{\alpha}{2\beta}\right).$$

On the other hand, a closed-form expression for the asymptotic IOWE of the 8-state convolutional encoder is not known. However, in [15], Sason *et al.* provided a numerical method for computing the asymptotic IOWE of convolutional encoders. Here, we use the numerical procedure proposed in [15] to compute the asymptotic IOWEs $a^{C_a}(\omega, \iota_a)$ and $a^{C_b}(\omega, 2\iota - \iota_a)$ in (7) to numerically compute the asymptotic spectral shape function.

The numerical evaluation of the asymptotic spectral shape function is displayed in Fig. 4 for 3D-TCs, $\lambda = 1$, and several code rates, and in Fig. 5 for $R = 1/3$ 3D-TCs and several values of $\lambda$. The behavior of the asymptotic spectral shape function is similar to the one of RMA codes [9], i.e., it is zero in the range $(0, \rho_0)$ and strictly positive for some $\rho > \rho_0$. Note that since closed-form expressions for $a^{C_a}(\omega, \iota_a)$ and $a^{C_b}(\omega, 2\iota - \iota_a)$ are not known, we cannot provide a formal proof that 3D-TCs are asymptotically good. However, the numerical evaluation of the asymptotic spectral shape function, together with extensive numerical experiments that show that $\Pr(d_{\min} < \lfloor \rho_0 N \rfloor) \longrightarrow 0$ as $N$ gets large, suggest that 3D-TCs for the parameters in Figs. 4 and 5 (for $\lambda$ down to $1/4$) are indeed asymptotically good. Furthermore, the results are in agreement with the finite-length analysis. On the other hand, for $\lambda = 1/8$, the asymptotic spectral shape function is strictly positive, meaning that the $d_{\min}$ does not grow linearly with the block length.

TABLE I
ESTIMATED GROWTH RATE COEFFICIENT $\rho_0$ OF 3D-TCS.

|  | $R = 1/3$ | $R = 1/2$ | $R = 2/3$ | $R = 4/5$ |
|---|---|---|---|---|
| $\lambda = 1$ | 0.102 | 0.077 | 0.052 | 0.030 |
| $\lambda = 1/2$ | 0.029 | 0.031 | 0.023 | 0.015 |
| $\lambda = 1/3$ | 0.008 | 0.010 | 0.008 | 0.006 |
| $\lambda = 1/4$ | 0.001 | 0.002 | 0.001 | 0.001 |

The estimated growth rate coefficient $\rho_0$ of 3D-TCs is reported in Table I for several code rates and values of the parameter $\lambda$. As expected, for a fixed code rate, the growth rate coefficient increases as $\lambda$ increases.

Notice that the growth rate coefficient for $R = 1/2$ is higher than for $R = 1/3$ for all values of $\lambda$, except for $\lambda = 1$. An heuristic explanation is that for $\lambda = 1/2$ and lower, to achieve $R = 1/2$, only parity bits sent directly to the channel are punctured. The minimum weight of the parity bits sent directly to channel will not grow linearly with the block length, since this would imply that the minimum distance of a conventional turbo code would grow linearly with the block length, which is not the case [6]. Thus, we would expect that there is not much difference in the growth rate between $R = 1/3$ and $R = 1/2$ for $\lambda \leq 1/2$, and the values for $R = 1/2$ can even be higher, since the rate is higher.

### B. Threshold Under Maximum-Likelihood (ML) Decoding

The asymptotic spectral shape function of a code ensemble can also be used to derive a threshold under ML decoding. An upper bound on the ML decoding threshold of a code ensemble, due to Divsalar [27], is given by

$$\left(\frac{E_b}{N_0}\right)_{\text{ML,threshold}} \leq \frac{1}{R} \cdot \max_{0 \leq \rho \leq 1} \left[\frac{(1 - e^{-2r(\rho)})(1 - \rho)}{2\rho}\right] \quad (8)$$

where $R$ is the code rate, $r(\rho)$ is the asymptotic spectral shape function, $E_b/N_0$ denotes the signal-to-noise ratio, and $(E_b/N_0)_{\text{ML,threshold}}$ is the ML decoding threshold. We computed the upper bound on the ML decoding threshold in (8) numerically for 3D-TC ensembles for several values of $\lambda$ and code rates. The results are given in Table II. For comparison purposes, we also report in the table the binary-input AWGN Shannon limit. For $\lambda = 1$ and $R = 1/3$, the ML decoding threshold is very close to the capacity limit, while the gap to capacity generally increases with the code rate. Also, note that similarly to the growth rate coefficient, the upper bounds on the ML decoding threshold decrease as the value of $\lambda$ increases. In Section VII, we will compare these upper bounds on the ML decoding threshold to iterative decoding thresholds computed from an EXIT chart analysis.

TABLE II
UPPER BOUNDS ON THE ML DECODING THRESHOLD OF 3D-TCS BASED ON DIVSALAR'S BOUND IN [27].

| | $R = 1/3$ | $R = 1/2$ | $R = 2/3$ | $R = 4/5$ |
|---|---|---|---|---|
| Capacity | -0.495 dB | 0.187 dB | 1.059 dB | 2.040 dB |
| $\lambda = 1$ | -0.440 dB | 0.319 dB | 1.325 dB | 2.475 dB |
| $\lambda = 1/2$ | -0.352 dB | 0.400 dB | 1.358 dB | 2.485 dB |
| $\lambda = 1/3$ | -0.272 dB | 0.514 dB | 1.452 dB | 2.509 dB |
| $\lambda = 1/4$ | -0.211 dB | 0.605 dB | 1.668 dB | 2.553 dB |

## V. QPPs OVER INTEGER RINGS

In the previous sections we analyzed the minimum distance properties of 3D-TC ensembles (generated by varying $\Pi$ and $\Pi_c$ over all possible permutations) and showed numerically that their $d_{\min}$ grows linearly with block length for certain values of $\lambda$. In the following, we consider the minimum distance properties of 3D-TCs with designed interleavers. In particular, we consider QPP interleavers. In this section, we establish notation and restate the criterion for existence of QPPs over integer rings. The interested reader is referred to [19, 28] for further details.

*Definition 1:* Given an integer $M \geq 2$, a polynomial $f(x) = f_1 x + f_2 x^2 \pmod{M}$, where $f_1$ and $f_2$ are nonnegative integers, is said to be a QPP over the ring of integers $\mathbb{Z}_M$ when $f(x)$ permutes $\{0, 1, 2, \ldots, M-1\}$.

In this paper, let the set of primes be $\mathcal{P} = \{2, 3, 5, 7, \ldots\}$. Then an integer $M$ can be factored as $M = \prod_{p \in \mathcal{P}} p^{n_{M,p}}$, where $n_{M,p} \geq 1$ for a finite number of $p$'s and $n_{M,p} = 0$ otherwise. For example, if $M = 3888 = 2^4 \times 3^5$ we have $n_{3888,2} = 4$ and $n_{3888,3} = 5$. For a quadratic polynomial $f(x) = f_1 x + f_2 x^2 \pmod{M}$, we will abuse the previous notation by writing $f_2 = \prod_{p \in \mathcal{P}} p^{n_{F,p}}$, i.e., the exponents of the prime factors of $f_2$ will be written as $n_{F,p}$ instead of the more cumbersome $n_{f_2,p}$ because we are interested in the factorization of $f_2$.

Let us denote $a$ divides $b$ by $a|b$ and by $a \nmid b$ otherwise. The greatest common divisor of $a$ and $b$ is denoted by $\gcd(a, b)$ and the least common multiple of $a$ and $b$ is denoted by $\text{lcm}(a, b)$. The necessary and sufficient condition for a quadratic polynomial $f(x)$ to be a PP is given below.

*Proposition 1 ([19, 28]):* Let $M = \prod_{p \in \mathcal{P}} p^{n_{M,p}}$. The necessary and sufficient condition for a quadratic polynomial $f(x) = f_1 x + f_2 x^2 \pmod{M}$ to be a PP can be divided into two cases.

1) Either $2 \nmid M$ or $4|M$ (i.e., $n_{M,2} \neq 1$)
   $\gcd(f_1, M) = 1$ and $f_2 = \prod_{p \in \mathcal{P}} p^{n_{F,p}}, n_{F,p} \geq 1, \forall p$ such that $n_{M,p} \geq 1$.
2) $2|M$ and $4 \nmid M$ (i.e., $n_{M,2} = 1$)
   $f_1 + f_2$ is odd, $\gcd(f_1, \frac{M}{2}) = 1$, and $f_2 = \prod_{p \in \mathcal{P}} p^{n_{F,p}}, n_{F,p} \geq 1, \forall p$ such that $p \neq 2$ and $n_{M,p} \geq 1$.

For example, if $M = 256$, then we determine from case 1) of Proposition 1 that $f_1 \in \{1, 3, 5, \ldots, 255\}$ (set of numbers relatively prime to $M$) and $f_2 \in \{2, 4, 6, \ldots, 254\}$ (set of numbers that contain 2 as a factor). This gives us $128 \times 127 = 16256$ possible pairs of coefficients $f_1$ and $f_2$ that make $f(x)$ a PP.

Finally, we remark that some QPPs have a quadratic inverse, i.e., the inverse permutation can also be generated by a QPP. We will not state the exact conditions here, but refer the interested reader to [28] for further details.

### A. Quasi-Cyclic Property of 3D-TCs Using QPPs

Assume tailbiting termination of the upper and lower constituent encoders and of the encoder in the patch of the 3D-TC. Furthermore, $1/\lambda$ is assumed to be a divisor of $K$, the length of the interleaver $\Pi$, and the puncturing pattern $\mathbf{p}$ is assumed to be regular, i.e., $\mathbf{p} = [1100 \cdots 00]$.

*Lemma 1:* The 3D-TC is quasi-cyclic with period $p$, where

$$p = l \cdot \text{lcm}(K/\gcd(2f_2, K), 1/\lambda, K/\gcd(2\tilde{f}_2, N_c)),$$

$f(x) = f_1 x + f_2 x^2 \pmod{K}$ and $\tilde{f}(x) = \tilde{f}_1 x + \tilde{f}_2 x^2 \pmod{N_c}$ generate the turbo code interleaver and the permutation in the patch, respectively, and $N_c$ is the input length to the patch, as defined in Section II. Furthermore, $l$ is the smallest positive integer solution to the quadratic congruence

$$2\lambda\tilde{p}((f_1 - 1)l + f_2\tilde{p}l^2) \equiv 0 \pmod{N_c} \quad (9)$$

where $\tilde{p} = p/l$.

*Proof:* Let $\mathbf{u}$ denote an input sequence to a 3D-TC, let $i$, $0 \leq i < K$, denote an arbitrary position in $\mathbf{u}$, and let $\overrightarrow{\mathbf{u}}$ denote a quasi-cyclic shift of period $p$ of $\mathbf{u}$. Now, the position $i + p \pmod{K}$ in $\overrightarrow{\mathbf{u}}$ is interleaved to $f(i+p) = f(i) + f(p) + 2f_2 ip \pmod{K}$. Furthermore, to make the difference between the interleaved positions $f(i + p)$ and $f(i)$ independent of $i$, or equivalently, $\mathbf{x}_b(\overrightarrow{\mathbf{u}})$ a quasi-cyclic shift of $\mathbf{x}_b(\mathbf{u})$, we require that $2f_2 p \equiv 0 \pmod{K}$, i.e., $p = l \cdot K/\gcd(2f_2, K)$, for some positive integer $l$. Also, to make $\mathbf{x}^p(\overrightarrow{\mathbf{u}})$ a quasi-cyclic shift of $\mathbf{x}^p(\mathbf{u})$, both $2\lambda(f(p) - p) \equiv 0 \pmod{N_c}$ (which gives (9)) and $(1/\lambda)|p$ must hold. Finally, to make $\mathbf{x}_c(\overrightarrow{\mathbf{u}})$ a quasi-cyclic shift of $\mathbf{x}_c(\mathbf{u})$, $2\lambda p$ needs to be a multiple of $N_c/\gcd(2\tilde{f}_2, N_c)$, or equivalently, $p$ needs to be a multiple of $K/\gcd(2\tilde{f}_2, N_c)$, since $\mathbf{x}^p(\overrightarrow{\mathbf{u}})$ is a quasi-cyclic shift of $\mathbf{x}^p(\mathbf{u})$ of period $2\lambda p$, and the result follows. ∎

Without the patch, the conventional binary turbo code is quasi-cyclic with period $K/\gcd(2f_2, K)$. We remark that a similar result will hold for 3D double-binary turbo codes [3] with *symbol interleaving* based on QPPs.

## VI. UPPER BOUNDS ON $d_{\min}$ WITH QPPS WITH A QUADRATIC INVERSE

In this section, we present upper bounds on $d_{\min}$ with QPPs with a quadratic inverse.

First, we state a general result from [29] on the minimum distance of a conventional binary turbo code with QPP interleavers. We assume tailbiting termination of the upper and lower constituent encoders.

*Theorem 1 ([29]):* The minimum distance of a conventional binary turbo code (of nominal rate $1/3$) using primitive feedback and monic feedforward polynomials of degree $\nu$ and QPPs with a quadratic inverse, is upper-bounded by $2(2^{\nu+1}+9)$.

We remark that Theorem 1 applies for all interleaver lengths $K$ and is achievable for a range of $K$-values [29]. The bound of Theorem 1 is due to an input-weight 6 codeword containing 3 input-weight 2 fundamental paths, or error events, in both the upper and lower constituent codewords. Also, the upper bound of Theorem 1 can be shown to hold with dual termination [30] as well, i.e., the upper and lower constituent encoders are forced to begin and end in the zero state, when $K \geq 2^{\nu+3} - 7$ [29].

### A. Upper Bounds on $d_{\min}$ With QPPs With a Quadratic Inverse for the 3D-TC With Any Patch

In this subsection, we consider the binary 3D-TC of Fig. 1 with any given patch and such that $4|(1/\lambda)|K$ and with regular puncturing pattern $\mathbf{p} = [1100\cdots 00]$. Note that it may be possible to derive bounds for other values of $\lambda$ using a similar procedure as the one outlined below. Also, we remark that the bounds here can be generalized to 3D-TCs with lower and upper constituent encoders other than the 8-state constituent encoder with feedforward polynomial $1+D+D^3$ and feedback polynomial $1 + D^2 + D^3$. However, in this work, we will constrain the analysis to this 8-state encoder and to the case when $4|(1/\lambda)|K$. On the other hand, we do not consider a specific encoder for $C_c$. In fact, in the analysis below, the only condition we require is that the encoder of the patch maps the all-zero sequence to the all-zero sequence. Also, we assume tailbiting termination of the upper and lower constituent encoders. Note that the termination method of the encoder in the patch is not an issue here.

The upper bounds in this section are based on certain critical codewords that always occur for some specific lengths for any QPP interleaver (with a quadratic inverse) for the conventional binary turbo code. To find these critical codewords, we have used the following strategy.

- First select a particular interleaver length $K$ and perform a random search for good pairs of QPPs using the triple impulse method [17] to estimate the $d_{\min}$ of the 3D-TC.
- Low-weight codewords identified by the triple impulse method are added to a list of codewords. Within this list of codewords, there are often codewords that give an all-zero input sequence into the patch, and it is sometimes possible to identify among these codewords certain types of codewords that occur repeatedly.

This is how we have found the critical codewords depicted in Figs. 10, 11, and 12 in Appendices A and B. These critical codewords give the following two theorems, where $f(x) = f_1 x + f_2 x^2 \pmod{K}$ is the QPP that generates the interleaver of the outer turbo code and $g(x) = g_1 x + g_2 x^2 \pmod{K}$ is its inverse.

*Theorem 2:* The minimum distance of a binary 3D-TC with feedforward polynomial $1 + D + D^3$ and feedback polynomial $1 + D^2 + D^3$ for the upper and lower constituent encoders and with QPP interleavers with a quadratic inverse, is upper-bounded by 67 when 1) the interleaver length $K$ satisfies the conditions

$$n_{K,p} \leq \begin{cases} 7, & \text{if } p = 2 \\ 1, & \text{otherwise,} \end{cases} \quad (10)$$

2) the encoder in the patch maps the all-zero sequence to the all-zero sequence, and 3) $4|(1/\lambda)|K$.

*Proof:* See Appendix A. ∎

For example, $K = 1504$ and $640$ satisfy the inequality in (10), since $1504 = 2^5 \times 47$ and $640 = 2^7 \times 5$.

*Theorem 3:* The minimum distance of a binary 3D-TC with feedforward polynomial $1 + D + D^3$ and feedback polynomial $1 + D^2 + D^3$ for the upper and lower constituent encoders and with QPP interleavers with a quadratic inverse, is upper-bounded by 27 (resp. 54) when 1) $2g_2 \equiv 0 \pmod{K}$ (resp. $4g_2 \equiv 0 \pmod{K}$), 2) the encoder in the patch maps the all-zero sequence to the all-zero sequence, and 3) $4|(1/\lambda)|K$.

*Proof:* See Appendix B. ∎

We remark that Theorem 3 can be formulated with the conditions $2f_2 \equiv 0 \pmod{K}$ and $4f_2 \equiv 0 \pmod{K}$ instead of the conditions $2g_2 \equiv 0 \pmod{K}$ and $4g_2 \equiv 0 \pmod{K}$ due to symmetry. Note that the condition $2g_2 \equiv 0 \pmod{K}$ is equivalent to the fact that the QPP $g(x) = g_1 x + g_2 x^2 \pmod{K}$ is indeed a linear PP, i.e., the same permutation is generated by a linear PP.

*Lemma 2:* If $K > 32$, then the upper bound on the $d_{\min}$ of 67 in Theorem 2 holds with dual termination as well.

*Proof:* See Appendix C. ∎

We remark that a lower bound on $K$ can be derived, in a similar fashion, to make the upper bounds on the $d_{\min}$ in Theorem 3 hold with dual termination as well. We omit the details for brevity.

Finally, note that in principle it is possible to derive bounds for other values of $\lambda$ and when puncturing is applied using the procedure above, both for binary and double-binary codes.

## VII. EXIT CHART ANALYSIS

In this section, we estimate the convergence thresholds of 3D-TCs through an EXIT chart analysis [31] on the AWGN channel using regular puncturing patterns $\mathbf{p}$. Also, higher code rates are obtained by randomly puncturing nonsystematic bits according to the puncturing strategy in Section II.

The EXIT charts of two 3D-TCs, with parameters $\lambda = 1/2$, $R = 1/3$, and $\lambda = 1/4$, $R = 2/3$, at an $E_b/N_0$ of $\gamma = 0.55$ dB and $\gamma = 1.72$ dB, respectively, are depicted in Fig. 6. The solid curves are the EXIT curves of the outer turbo code, while the dashed curves are the EXIT curves of the inner encoder. In both cases a tunnel between the two EXIT curves is observed, meaning that convergence is possible at this $E_b/N_0$. Note that both EXIT functions of the inner code and the outer turbo code depend on the $E_b/N_0$. For each pair of curves, a vertical step between the lower curve (i.e., the EXIT curve of the turbo code) and the upper curve (i.e., the EXIT curve of the inner code) represents a single activation of decoder $C_c^{-1}$, while a horizontal step between the upper curve and the lower curve represents an unspecified number of activations of decoders $C_a^{-1}$ and $C_b^{-1}$ until nothing more can be gained. For $\lambda = 1/2$, $R = 1/3$, the tunnel between the two curves opens at $\gamma = 0.52$ dB, predicting a convergence threshold around this value. For $\lambda = 1/4$, $R = 2/3$, the tunnel opens at $\gamma = 1.68$ dB. The convergence thresholds of 3D-TCs are given in Table III for several code rates and values of $\lambda$. The convergence thresholds were computed assuming a block length $K = 10^6$ bits[1]. For a given code rate, the best convergence threshold is achieved by the fully parallel concatenated code ($\lambda = 0$), while increasing $\lambda$ leads to poorer thresholds. On the other hand, in terms of $d_{\min}$, the behavior is the opposite, i.e., larger minimum distances are obtained by increasing $\lambda$. This suggests that there is a trade-off between iterative convergence threshold and $d_{\min}$ growth rate.

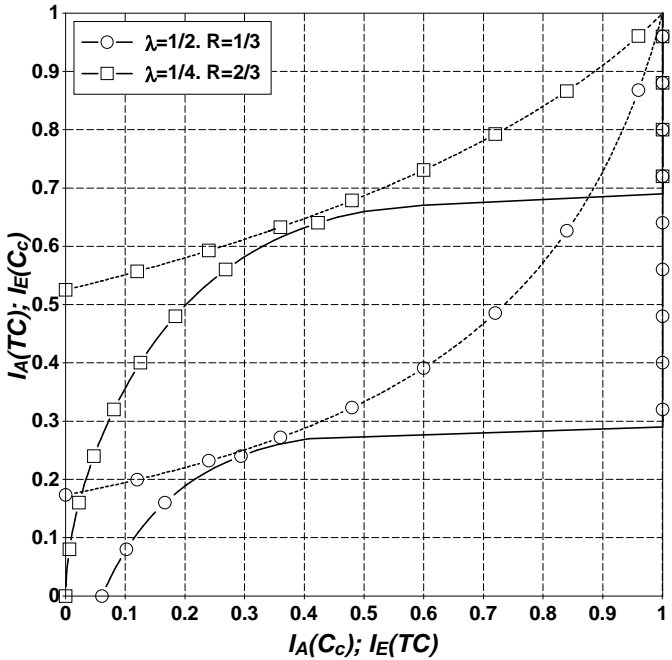

Fig. 6. EXIT charts of $\lambda = 1/2$, $R = 1/3$, 3D-TC ($\gamma = 0.55$ dB) and of $\lambda = 1/4$, $R = 2/3$, 3D-TC ($\gamma = 1.72$ dB). The dashed curves are the EXIT curves of the inner encoder, while the solid curves are the EXIT curves of the outer turbo code.

TABLE III
CONVERGENCE THRESHOLDS OF 3D-TCs.

|  | $R = 1/3$ | $R = 1/2$ | $R = 2/3$ | $R = 4/5$ |
|---|---|---|---|---|
| $\lambda = 1$ | 1.26 dB | 2.00 dB | 2.99 dB | 4.14 dB |
| $\lambda = 1/2$ | 0.52 dB | 1.24 dB | 2.23 dB | 3.39 dB |
| $\lambda = 1/3$ | 0.30 dB | 1.00 dB | 1.88 dB | 3.00 dB |
| $\lambda = 1/4$ | 0.20 dB | 0.85 dB | 1.68 dB | 2.79 dB |
| $\lambda = 0$ | -0.05 dB | 0.62 dB | 1.53 dB | 2.46 dB |

[1]We remark that the thresholds in Table III are slightly different from the thresholds in Table II in [32]. This is due to the fact that the thresholds in [32] were computed using a shorter block length.

It is interesting to note that the iterative decoding thresholds get worse for increasing values of $\lambda$, while the upper bounds on the ML decoding threshold (see Table II) improve. This behavior can be explained from the fact that the latter depends on the code while the former depends also on the decoding algorithm. The sub-optimality of the iterative decoding algorithm is expected to be higher when the contribution of the patch is higher (i.e., for increasing values of $\lambda$), which explains the results in Table III compared to the upper bounds on the ML decoding threshold in Table II.

## VIII. NUMERICAL RESULTS

In this section, we present some numerical results from a random search for pairs of QPPs with $\lambda = 1/4$ and with regular puncturing pattern $\mathbf{p} = [11000000]$, which give good estimated minimum distance for the binary rate-$1/3$ 3D-TC. As shown above, $\lambda = 1/4$ is a good compromise between minimum distance and convergence threshold. To estimate the minimum distance, we used the triple impulse method [17]. Also, to speed up the search, we limited the search space through the conditions $2^l f_2 \not\equiv 0 \pmod{K}$, $l = 2$ and $4$, whenever appropriate. The rationale behind the condition $16 f_2 \not\equiv 0 \pmod{K}$ can be found in the proof of Theorem 2 in Appendix A. In the search, all three constituent encoders were forced to begin and end in the zero state. The results, for some specific lengths and rates (high rates are obtained by puncturing nonsystematic bits, as explained in Section II), are tabulated in Table IV. For each code rate, the minimum distances of the 3D-TCs, estimated using the triple impulse method, are denoted by $\hat{d}_{\min}^{(R)}$, where $R$ is the code rate. For comparison purposes, we have also tabulated the probabilistic lower bound on the $d_{\min}$ from (4) with $\epsilon = 0.5$, denoted by $d_{\min,\text{LB}}^{(R)}$, using random puncturing patterns.

For rate $1/3$, the results are reported in column 7 in Table IV. The $K$-values in bold indicate lengths where the upper bound in Theorem 2 appears to be tight. In the table, we also report (in column 8) the probabilistic lower bound on the $d_{\min}$ from (4), and the optimum $d_{\min}$ (in column 15) for a conventional rate-$1/3$ binary turbo code with a QPP interleaver, denoted by $d_{\min}^{\text{TC},(1/3)}$. As expected, the designed QPP interleavers improve the $d_{\min}$ significantly with respect to the probabilistic lower bound. Moreover, the 3D-TCs achieve much higher minimum distances than the best conventional turbo codes. Therefore, they are expected to exhibit much lower error floors.

The results for rates $1/2$, $2/3$, and $4/5$ are tabulated in columns 9, 11, and 13 in Table IV, respectively. Specific periodic puncturing patterns have been found by computer search, following the puncturing strategy in Section II. In more detail, for rates less than $2/3$, only bits from $\mathbf{x}^{\text{ch}}$ are punctured. In particular, for rate $R = 2/3$, all bits from $\mathbf{x}^{\text{ch}}$ are punctured. For rates larger than $2/3$, all bits from $\mathbf{x}^{\text{ch}}$ and some bits from $\mathbf{x}_c$ are punctured.

In the computer search, both error rate performance in the error floor region (i.e., the minimum distance) and the waterfall region have been considered. Performance in the waterfall region is particularly important for rate $4/5$, since some puncturing patterns show very poor error rate performance. In fact, some puncturing patterns do not give a distinguishable waterfall region at all. This is the case for 54 of the 70 puncturing patterns of period 8 for $\mathbf{x}_c$. In more detail, for rate $1/2$, we have looked at all 15 ($\binom{6}{2}$) periodic puncturing patterns of period 6 for $\mathbf{x}_a^{\text{ch}}$ and $\mathbf{x}_b^{\text{ch}}$, and all their combinations, i.e., we have looked at $15 \times 15 = 225$ periodic puncturing patterns of period 12 for $\mathbf{x}^{\text{ch}}$. For rate $2/3$, according to the puncturing strategy above, there is only one puncturing pattern to check, and finally, for rate $4/5$, we have looked at all 70 ($\binom{8}{4}$) periodic puncturing patterns of period 8 for $\mathbf{x}_c$. In the computer search, for each candidate puncturing pattern, an upper bound on the $d_{\min}$ was computed using the triple impulse method, which generated an ordered list (the candidates were ordered according to the computed upper bound on the $d_{\min}$) of puncturing patterns. Also, to speed up the search, candidates where rejected if a codeword of weight less than some *running* $d_{\min}$ value was found. Finally, an estimate of the $d_{\min}$ of the best candidates in the ordered list, in the order of the list, was computed using a *stronger* version of the triple impulse method. By a stronger version of the triple impulse method we mean an impulse method where the impulse ranges have been extended quite a bit. The actual fine-tuning of the impulse ranges has been done based on experimental results and running time considerations. For rate $4/5$, an initial filtering of the candidate puncturing patterns based on convergence can also be done.

TABLE IV
RESULTS FROM A RANDOM SEARCH FOR PAIRS OF QPPS WITH $\lambda = 1/4$, BOTH WITH A QUADRATIC INVERSE, IN WHICH THE FIRST QPP $f(x)$ GENERATES THE TURBO CODE INTERLEAVER AND THE SECOND QPP $\tilde{f}(x)$ GENERATES THE PERMUTATION IN THE PATCH. HIGHER RATES ARE OBTAINED BY PUNCTURING THE NONSYSTEMATIC BITS USING OPTIMIZED (BY COMPUTER SEARCH) PERIODIC PUNCTURING PATTERNS.

| $K$ | $f_1$ | $f_2$ | $N_c$ | $\tilde{f}_1$ | $\tilde{f}_2$ | $\hat{d}_{\min}^{(1/3)}$ | $d_{\min,\text{LB}}^{(1/3)}$ | $\hat{d}_{\min}^{(1/2)}$ | $d_{\min,\text{LB}}^{(1/2)}$ | $\hat{d}_{\min}^{(2/3)}$ | $d_{\min,\text{LB}}^{(2/3)}$ | $\hat{d}_{\min}^{(4/5)}$ | $d_{\min,\text{LB}}^{(4/5)}$ | $d_{\min}^{\text{TC},(1/3)}$ |
|---|---|---|---|---|---|---|---|---|---|---|---|---|---|---|
| 512 | 175 | 192 | 256 | 15 | 192 | 67 | 33 | 38 | 19 | 17 | 11 | 15 | 7 | 38 [20] |
| **640** | 631 | 40 | 320 | 21 | 180 | 67 | 36 | 41 | 21 | 17 | 12 | 16 | 8 | 39 [29] |
| 768 | 613 | 24 | 384 | 73 | 216 | 79 | 39 | 42 | 24 | 22 | 14 | 17 | 9 | 39 [29] |
| 1024 | 465 | 224 | 512 | 157 | 160 | 93 | 45 | 54 | 28 | 23 | 17 | 22 | 11 | 45 [20] |
| **1504** | 299 | 188 | 752 | 147 | 282 | 67 | 57 | 41 | 36 | 17 | 22 | 16 | 15 | 50 [29] |
| 2048 | 673 | 448 | 1024 | 71 | 192 | 147 | 69 | 110 | 45 | 64 | 29 | 40 | 19 | 50 [29] |

For comparison purposes, we report, in columns 10, 12, and 14 in Table IV, the probabilistic lower bound (with $\epsilon = 0.5$) on the $d_{\min}$ from (4) using random puncturing patterns for rates $1/2$, $2/3$, and $4/5$, respectively. Note that the designed QPP interleavers improve the $d_{\min}$ with respect to the probabilistic lower bound, even for high rates, except for $K = 1504$ and rate $2/3$. However, $K = 1504$ is indicated in bold in Table IV, which means that the upper bound in Theorem 2 applies. Thus, this is not a good value for $K$ with QPP interleavers, and should therefore be avoided.

### A. Simulation Results

In Fig. 7, we give FER curves on the AWGN channel of a nominal rate-$1/3$ binary 3D-TC with $K = 1024$ and 2048. The turbo code interleavers were generated by the QPPs $f(x) = 465x + 224x^2 \pmod{1024}$ and $f(x) = 673x + 448x^2$

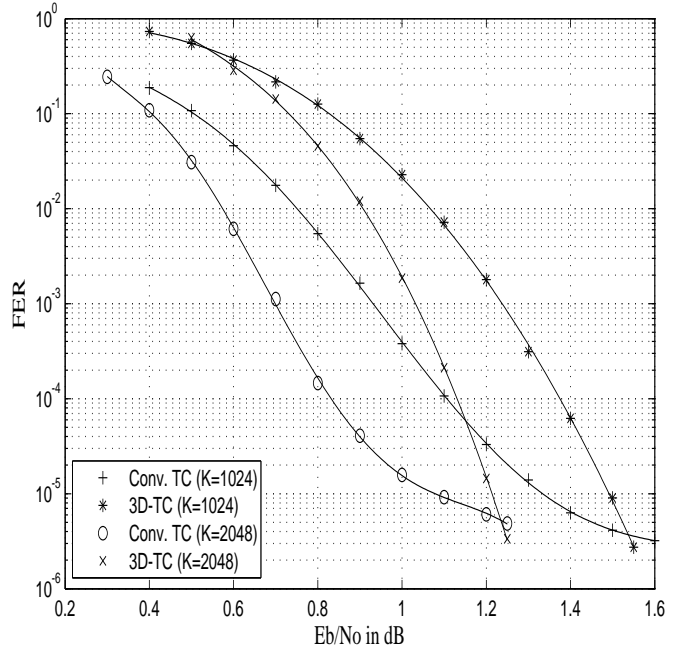

Fig. 7. *Comparison of FER performance of iterative decoding of conventional turbo codes and binary 3D-TCs. The nominal code rate is $1/3$.*

$\pmod{2048}$, and the permutations in the patch were generated by the QPPs $\tilde{f}(x) = 157x + 160x^2 \pmod{512}$ and $\tilde{f}(x) = 71x + 192x^2 \pmod{1024}$, for $K = 1024$ and 2048, respectively (see Table IV). In the simulation, we used $\lambda = 1/4$ and a maximum of 16 iterations. The actual code rate is $(K - 2\nu - \tilde{\nu})/3K$ due to trellis termination, where $\nu$ is the constraint length of the upper and lower constituent encoders and $\tilde{\nu}$ is the constraint length of the third encoder in the patch. Furthermore, to simplify the decoders, the max-log approximation with scaling of the extrinsic values was used within the Bahl-Cocke-Jelinek-Raviv (BCJR) algorithm. For comparison purposes, the performance of conventional binary turbo codes with input length $K = 1024$ and 2048, and with rate $(K - 2\nu)/3K$ (due to dual termination), is also given in the figure. The turbo code interleaver for $K = 2048$ was generated by the QPP $f(x) = 21x + 128x^2 \pmod{2048}$, which gives the optimum (the upper bound in Theorem 1 is achieved) minimum distance of 50 [29], and the same maximum number of decoding iterations was used. For $K = 1024$, the turbo code interleaver was generated by the QPP $f(x) = 245x + 448x^2$ (mod 1024), which is a $d_{\min}$-optimum QPP interleaver giving a $d_{\min}$ of 45 [29]. As can be seen from Fig. 7, there is a performance loss of about 0.3 dB in the waterfall region for the 3D-TC compared to the conventional turbo code. This loss in performance is consistent with the EXIT chart analysis of the previous section (see Table III). However, for very low error rates, the performance of the 3D-TC will clearly be superior to the performance of the conventional turbo code, due to a much larger $d_{\min}$.

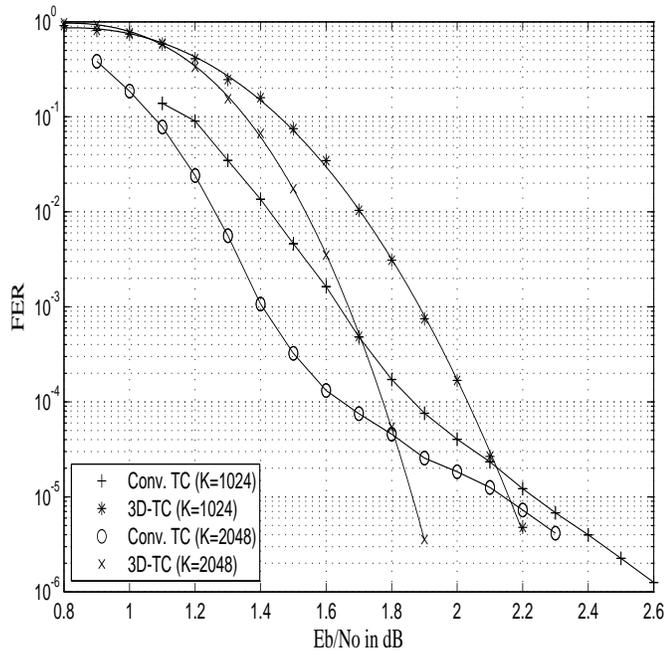

Fig. 8. *Comparison of FER performance of iterative decoding of conventional turbo codes and binary 3D-TCs. The nominal code rate is* $1/2$.

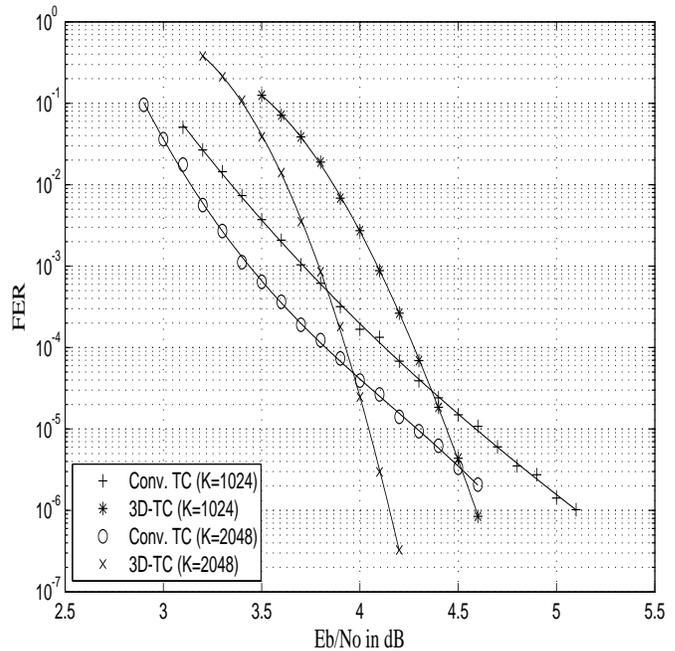

Fig. 9. *Comparison of FER performance of iterative decoding of conventional turbo codes and binary 3D-TCs. The nominal code rate is* $4/5$.

In Figs. 8 and 9, we give the FER performance of the 3D-TCs from Fig. 7 with different puncturing patterns. For comparison purposes, the performance of the conventional binary turbo codes from Fig. 7 with different puncturing patterns is also plotted. The nominal code rate of the codes in Fig. 8 is $1/2$, while the nominal code rate of the codes in Fig. 9 is $4/5$. In both cases, optimized puncturing patterns (the puncturing patterns used to produce the entries in Table IV) were used in the simulations of the 3D-TCs. Thus, the 3D-TCs in Fig. 8 have estimated minimum distances of 54 and 110 for $K = 1024$ and 2048, respectively. In Fig. 9, the 3D-TCs have estimated minimum distances of 22 and 40 for $K = 1024$ and 2048, respectively. Also, for a fair comparison, the puncturing patterns for the conventional binary turbo codes were optimized as well, using the same interleavers as for the rate-$1/3$ mother codes, which were optimized through a QPP interleaver search, giving the minimum distance values in column 15 of Table IV. In fact, the conventional binary turbo codes in Fig. 8 have minimum distances of 24 and 25 for $K = 1024$ and 2048, respectively. In Fig. 9, the conventional binary turbo codes have minimum distances of 8 and 9 for $K = 1024$ and 2048, respectively. From the figures, we observe no error floor down to a FER of about $10^{-6}$ for the 3D-TCs. Also, as expected, there is a loss in performance in the waterfall region with respect to conventional binary turbo codes. However, for low error rates, the performance of the 3D-TCs will clearly be superior to the performance of the conventional binary turbo codes for both rates. The conventional turbo code shows a flattening between a FER of $10^{-4} - 10^{-5}$ and $10^{-3} - 10^{-4}$ for $R = 1/2$ and $4/5$, respectively, while significantly lower error floors are expected for the 3D-TC due to its better minimum distance.

## IX. CONCLUSION AND FUTURE WORK

In this work, we presented a finite-length and an asymptotic minimum distance analysis of 3D-TCs with binary constituent encoders. By using a numerical procedure, recently proposed by Sason *et al.*, to compute the asymptotic IOWE of convolutional encoders, we numerically evaluated the asymptotic spectral shape function of 3D-TCs and showed (numerically) that for certain parameters, the 3D-TC ensemble is asymptotically good, in the sense that its typical minimum distance asymptotically grows linearly with the block length. The results were supported by the finite-length analysis. Higher rates obtained through random puncturing were also considered. In the second part of the paper, designed QPP interleavers were considered. In particular, we derived some useful upper bounds on the $d_{\min}$ when using this type of interleaver with a quadratic inverse. A random search for pairs of QPPs for use in the 3D-TC was performed, and the best codes (in terms of estimated minimum distance) were compared to a probabilistic lower bound on the $d_{\min}$. Higher rates were obtained through specific optimized (by computer search) periodic puncturing patterns. This comparison showed that the use of designed QPP interleavers can improve the minimum distance significantly. Finally, an EXIT chart analysis was conducted, iterative decoding thresholds and upper bounds on the ML decoding threshold were computed, and some simulation results were presented to compare the error rate performance of the 3D-TC to that of the conventional turbo code.

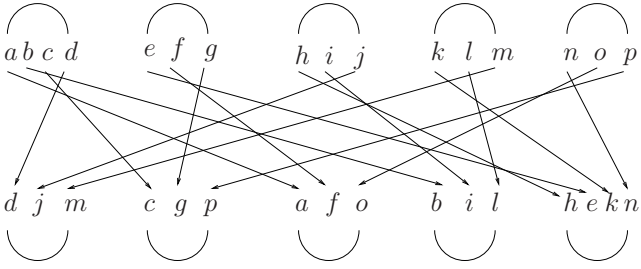

Fig. 10. An input-weight 16 critical codeword with QPP interleavers with a quadratic inverse for a conventional binary turbo code without the patch.

An interesting topic for future work is to derive upper bounds on the $d_{\min}$ for 3D-TCs with QPP interleavers with a quadratic inverse for other values of the interleaver length $K$, i.e., for interleaver lengths that do not satisfy the constraints of Theorem 2, for instance, when $K$ is a power of two. Also, as shown by Theorem 1, there is an upper bound on the $d_{\min}$ which is independent of the interleaver length $K$ when using QPP interleavers with a quadratic inverse for a conventional binary turbo code. Can a similar bound be found for 3D-TCs?

## APPENDIX A
## PROOF OF THEOREM 2

In Fig. 10, an input-weight 16 critical codeword is shown for the conventional binary turbo code. The upper and lower constituent codewords contain 5 fundamental paths each. The interleaving of the systematic 1-positions is indicated by arrows and the 16 first letters of the English alphabet. The fundamental paths are of the following three types

1) $0 \xrightarrow{1/11} 3 \xrightarrow{1/10} 4 \xrightarrow{0/00} 2 \xrightarrow{0/00} 1 \xrightarrow{0/01} 6 \xrightarrow{1/11} 0$
2) $0 \xrightarrow{1/11} 3 \xrightarrow{0/01} 7 \xrightarrow{1/10} 6 \xrightarrow{0/00} 3 \xrightarrow{0/01} 7 \xrightarrow{0/01} 5 \xrightarrow{0/01}$
   $4 \xrightarrow{0/00} 2 \xrightarrow{0/00} 1 \xrightarrow{0/01} 6 \xrightarrow{1/11} 0$ (11)
3) $0 \xrightarrow{1/11} 3 \xrightarrow{1/10} 4 \xrightarrow{0/00} 2 \xrightarrow{0/00} 1 \xrightarrow{1/10} 5 \xrightarrow{0/01} 4 \xrightarrow{0/00}$
   $2 \xrightarrow{0/00} 1 \xrightarrow{0/01} 6 \xrightarrow{1/11} 0$

where a state transition from state $a$ to state $b$ with input label $x$ and output label $yz$ is denoted by $a \xrightarrow{x/yz} b$. To be more specific, for the upper constituent codeword, the first (from the left) fundamental path is of type 3), the second is of type 1), the third is of type 2), the fourth is of type 2), and the fifth is of type 1). For the lower constituent codeword, the first fundamental path is of type 1), the second is of type 2), the third is of type 2), the fourth is of type 1), and the fifth is of type 3). The overall weight of the codeword is at most 64. It is an upper bound, since some of the fundamental paths may overlap. To make the structure in Fig. 10 an actual codeword, the following conditions must hold

a) $f(x+9) + 1 \equiv f(g(f(x+1)+1)+8) \pmod{K}$
b) $f(x+9) + 5 \equiv f(g(f(x+1)+5)+8) \pmod{K}$
c) $f(x+4) + 2 \equiv f(g(f(x)+2)+4) \pmod{K}$
d) $f(x+4) + 10 \equiv f(g(f(x)+10)+4) \pmod{K}$
e) $f(g(f(x+1)+1)-2)+1 \equiv f(g(f(x)+2)-1) \pmod{K}$
f) $f(g(f(x+1)+1)-2)+4 \equiv f(g(f(x+1)+5)-2) \pmod{K}$
g) $f(g(f(x+1)+1)-2)+9 \equiv f(g(f(x)+10)-1) \pmod{K}$

where $x \in \mathbb{Z}_K$ is the leftmost systematic 1-position in the upper constituent codeword, $g(x) = g_1 x + g_2 x^2 \pmod{K}$ is the inverse of $f(x) = f_1 x + f_2 x^2 \pmod{K}$, and $f(x)$ is the QPP that generates the turbo code interleaver. These congruences reduce to

a) $16 f_2 g_2 (1 + 2f(x+1)) + 16 f_2 g_1 \equiv 0 \pmod{K}$
b) $80 f_2 g_2 (5 + 2f(x+1)) + 80 f_2 g_1 \equiv 0 \pmod{K}$
c) $32 f_2 g_2 (1 + f(x)) + 16 f_2 g_1 \equiv 0 \pmod{K}$
d) $160 f_2 g_2 (5 + f(x)) + 80 f_2 g_1 \equiv 0 \pmod{K}$
e) $4 f_2 g_2 (-1 + 2 f_1 + 2 f_2 + 4 f_2 x) \equiv 0 \pmod{K}$
f) $32 f_2 g_2 (3 + f(x+1)) + 16 f_2 g_1 \equiv 0 \pmod{K}$
g) $16 f_2 g_1 + 4 f_2 g_2 (49 + 2(4 f(x) - f_1 - f_2 - 2 f_2 x)) \equiv 0 \pmod{K}$

If $27 \nmid K$, then these congruences reduce further to

a) $16 f_2 \equiv 0 \pmod{K}$
b) $80 f_2 \equiv 0 \pmod{K}$
c) $16 f_2 \equiv 0 \pmod{K}$
d) $80 f_2 \equiv 0 \pmod{K}$
e) $0 \equiv 0 \pmod{K}$
f) $16 f_2 \equiv 0 \pmod{K}$
g) $16 f_2 \equiv 0 \pmod{K}$

since $4 f_2 g_2 \equiv 0 \pmod{K}$ [28, Theorem 3.5] when $27 \nmid K$, and $g_1$ is relatively prime with $K$, since $4|K$. Now, if $16 f_2 \equiv 0 \pmod{K}$, then all these congruences are satisfied, from which it follows that $n_{F,2} + 4 \geq n_{K,2}$. Using [28, Theorem 3.6], it follows that the inequality above is true for all values of $f_2$ if

$$n_{K,2} \leq \begin{cases} 4 + \max\left(\left\lceil \frac{n_{K,2}-2}{2} \right\rceil, 1\right), & \text{if } n_{K,2} > 1 \\ 4, & \text{if } n_{K,2} = 0, 1 \end{cases}$$

$$n_{K,3} \leq \begin{cases} \max\left(\left\lceil \frac{n_{K,3}-1}{2} \right\rceil, 1\right), & \text{if } n_{K,3} > 0 \\ 0, & \text{if } n_{K,3} = 0 \end{cases}$$

$$n_{K,p} \leq \left\lceil \frac{n_{K,p}}{2} \right\rceil, \quad \text{if } p \neq 2, 3$$

which reduces to

$$n_{K,p} \leq \begin{cases} 7, & \text{if } p = 2 \\ 1, & \text{otherwise} \end{cases} \quad (12)$$

which is one of the conditions stated in the theorem. We will now look at the first 1-position for each of the upper fundamental paths in Fig. 10. Note that since $4|K$ (the third assumption in the theorem), we can change the order of $\pmod{K}$ and $\pmod 4$, i.e., $(x \pmod K) \pmod 4 = x \pmod 4$ for any integer $x$. Thus, the 1-positions are $x$, $g(f(x)+2)-1 \equiv x+2g_1-1 \pmod 4$, $g(f(x+1)+1)-2 \equiv x-1+g_1+g_2+2g_2 f(x+1) \pmod 4$, $g(f(x+1)+5)-2 \equiv x-1+g_1+g_2+2g_2 f(x+1) \pmod 4$, and $g(f(x)+10)-1 \equiv x+2g_1-1 \pmod 4$. With $x \equiv 1$ (mod 4), we get the 1-positions 1 (mod 4), $2g_1$ (mod 4), $g_1 + g_2$ (mod 4), $g_1 + g_2$ (mod 4), and $2g_1$ (mod 4).

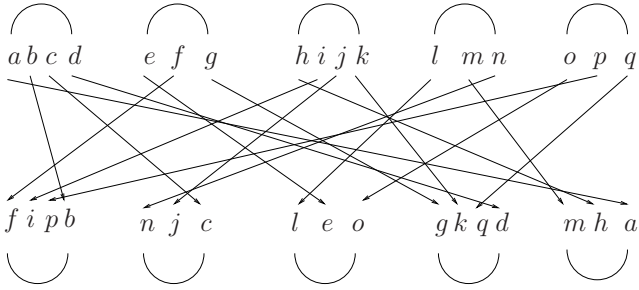

Fig. 11. An input-weight 17 critical codeword with QPP interleavers with a quadratic inverse for a conventional binary turbo code without the patch.

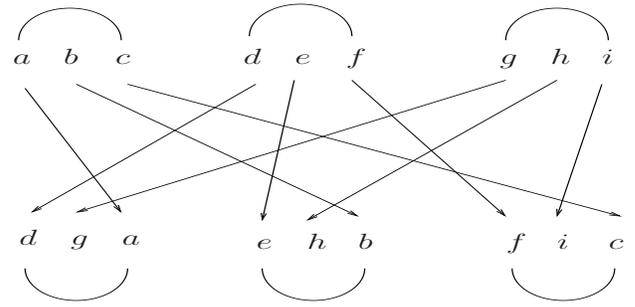

Fig. 12. An input-weight 9 critical codeword with QPP interleavers with a quadratic inverse for a conventional binary turbo code without the patch.

For the second component we get the following. The 1-positions are $f(x+9) \equiv f(x+1) \pmod 4$, $f(x+4) \equiv f(x) \pmod 4$, $f(x)$, $f(x+1)$, and $f(g(f(x+1)+1)-2) \equiv f(x) - f_1 + f_2 + 1 + 2f_2 x \pmod 4$. With $x \equiv 1 \pmod 4$, we get the 1-positions $2f_1 \pmod 4$, $f_1 + f_2 \pmod 4$, $f_1 + f_2 \pmod 4$, $2f_1 \pmod 4$, and $1 \pmod 4$.

Note that $g_1 + g_2 \equiv f_1 + f_2 \pmod 4$ and $2f_1 \equiv 2g_1 \equiv 2 \pmod 4$ when $4 \mid K$.

From the fundamental paths in (11), it follows that the input to the third encoder within the patch is the all-zero sequence when $g_1 + g_2 \equiv f_1 + f_2 \equiv 1 \pmod 4$. Note that the other case, i.e., when $g_1 + g_2 \equiv f_1 + f_2 \equiv 3 \pmod 4$ ($f_1 + f_2$ and $g_1 + g_2$ are odd) does *not* give an all-zero input sequence.

In Fig. 11, an input-weight 17 critical codeword is shown for the conventional binary turbo code. The upper and lower constituent codewords contain 5 fundamental paths each. The interleaving of the systematic 1-positions is indicated by arrows and the 17 first letters of the English alphabet. The fundamental paths are of the same type as those in Fig. 10. To be more specific, for the upper constituent codeword, the first (from the left) fundamental path is of type 3), the second is of type 2), the third is of type 3), the fourth is of type 1), and the fifth is of type 2). For the lower constituent codeword, the first fundamental path is of type 3), the second is of type 2), the third is of type 1), the fourth is of type 3), and the fifth is of type 2). The overall weight of the codeword is at most 67. To make the structure in Fig. 11 an actual codeword, the following conditions must hold

a) $g(f(x) - 8) + 4 \equiv g(f(x+4) - 8) \pmod K$
b) $g(f(x) - 8) + 1 \equiv g(f(x+1) - 8) \pmod K$
c) $g(f(x) - 8) + 9 \equiv g(f(x+9) - 8) \pmod K$
d) $g(f(x+9) - 9) - 8 \equiv g(f(x+1) - 9) \pmod K$
e) $g(f(x+9) - 9) - 10 \equiv g(f(g(f(x) - 10) - 1) + 1) \pmod K$
f) $g(f(x+4) - 10) \equiv g(f(x) - 10) + 4 \pmod K$
g) $f(g(f(x+1) - 5) - 2) \equiv f(g(f(x) - 10) - 1) + 5 \pmod K$
h) $g(f(x+9) - 5) - 8 \equiv g(f(x+1) - 5) \pmod K$

where $x \in \mathbb{Z}_K$ is the leftmost systematic 1-position in the upper constituent codeword. We can show that if $16g_2 \equiv 0 \pmod K$ and $27 \nmid K$, then all these congruences are satisfied for any $x \in \mathbb{Z}_K$, and it follows that $n_{G,2} + 4 \geq n_{K,2}$. Using [28, Theorem 3.6], we get the same conditions as in (12), which is one of the conditions stated in the theorem.

We will now look at the first 1-position for each of the upper fundamental paths in Fig. 11. Note that since $4 \mid K$ (the third assumption in the theorem), we can change the order of (mod $K$) and (mod 4). Thus, the 1-positions are $x$, $g(f(x+9) - 9) - 10 \equiv x - 1 - g_1 + g_2 - 2g_2 f(x+1) \pmod 4$, $g(f(x) - 8) \equiv x \pmod 4$, $g(f(x) - 10) - 1 \equiv x - 2g_1 - 1 \pmod 4$, and $g(f(x+1) - 5) - 2 \equiv x - 1 - g_1 + g_2 - 2g_2 f(x+1) \pmod 4$. With $x \equiv 1 \pmod 4$, we get the 1-positions 1 (mod 4), $3g_1 + g_2$ (mod 4), 1 (mod 4), $2g_1$ (mod 4), and $3g_1 + g_2$ (mod 4).

For the second component we get the following. The 1-positions are $f(x+1) - 9 \equiv f(x+1) - 1 \pmod 4$, $f(x+4) - 10 \equiv f(x) - 2 \pmod 4$, $f(g(f(x) - 10) - 1) \equiv f(x) - 2 - f_1 + f_2 - 2f_2 x \pmod 4$, $f(x+9) - 9 \equiv f(x+1) - 1 \pmod 4$, and $f(x) - 10 \equiv f(x) - 2 \pmod 4$. With $x \equiv 1 \pmod 4$, we get the 1-positions $2f_1 - 1 \pmod 4$, $f_1 + f_2 - 2 \pmod 4$, 2 (mod 4), $2f_1 - 1 \pmod 4$, and $f_1 + f_2 - 2 \pmod 4$.

From the fundamental paths in (11), it follows that the input to the third encoder within the patch is the all-zero sequence when $g_1 + g_2 \equiv f_1 + f_2 \equiv 3 \pmod 4$. Note that the other case, i.e., when $g_1 + g_2 \equiv f_1 + f_2 \equiv 1 \pmod 4$, does *not* give an all-zero input sequence.

The result of Theorem 2 follows by combining the results above, i.e., there is an upper bound of 64 (resp. 67) when $g_1 + g_2 \equiv f_1 + f_2 \equiv 1 \pmod 4$ (resp. $g_1 + g_2 \equiv f_1 + f_2 \equiv 3 \pmod 4$).

## APPENDIX B
## PROOF OF THEOREM 3

In Fig. 12, an input-weight 9 critical codeword is shown for the conventional binary turbo code. The upper and lower constituent codewords contain 3 fundamental paths each. The interleaving of the systematic 1-positions is indicated by arrows and the 9 first letters of the English alphabet. The fundamental paths are all of type 1), and the overall weight of the codeword is at most 27. To make the structure in Fig. 12 an actual codeword, the following conditions must hold

a) $g(f(x) - 5) + 1 \equiv g(f(x+1) - 5) \pmod K$
b) $g(f(x) - 5) + 5 \equiv g(f(x+5) - 5) \pmod K$
c) $g(f(x) - 4) + 1 \equiv g(f(x+1) - 4) \pmod K$
d) $g(f(x) - 4) + 5 \equiv g(f(x+5) - 4) \pmod K$

where $x \in \mathbb{Z}_K$ is the leftmost systematic 1-position in the upper constituent codeword. We can show that if $2g_2 \equiv 0$

(mod $K$), then all these congruences are satisfied. Furthermore, the input to the third encoder within the patch is the all-zero sequence. Thus, if $2g_2 \equiv 0 \pmod{K}$, then there is an upper bound of 27 on the minimum distance $d_{\min}$.

Reasoning in a similar manner with an input-weight 18 codeword, we can show that the minimum distance $d_{\min}$ is upper-bounded by 54 if $4g_2 \equiv 0 \pmod{K}$. The details are omitted for brevity.

# APPENDIX C
## PROOF OF LEMMA 2

To prove the lemma, we need the following result.

*Lemma 3:* If $4|K$ and $x_1 \equiv x_2 \pmod 4$, then $f(x_1) \equiv f(x_2) \pmod 4$ for all $x_1, x_2 \in \mathbb{Z}_K$ and all QPPs $f(x) = f_1 x + f_2 x^2 \pmod{K}$.

*Proof:* Since $4|K$, we can change the order of $\pmod{K}$ and $\pmod 4$, i.e., $(x \pmod{K}) \pmod 4 = x \pmod 4$ for any integer $x$. Thus,

- if $x \equiv 0 \pmod 4$, then $f(x) = 4\tilde{x}f_1 + 16\tilde{x}^2 f_2 \pmod{K} \equiv 0 \pmod 4$, for some integer $\tilde{x}$.
- If $x \equiv 1 \pmod 4$, then $f(x) = (4\tilde{x}+1)f_1 + (4\tilde{x}+1)^2 f_2 \pmod{K} \equiv f_1 + f_2 \pmod 4$, for some integer $\tilde{x}$. Since $f_2$ is even and $f_1$ is odd (see item 1) of Definition 1), $f(x) \pmod 4$ is odd.
- If $x \equiv 2 \pmod 4$, then $f(x) = (4\tilde{x}+2)f_1 + (4\tilde{x}+2)^2 f_2 \pmod{K} \equiv 2f_1 \pmod 4$, for some integer $\tilde{x}$. Furthermore, since $f_1$ is odd (see item 1) of Definition 1), $f(x) \equiv 2 \pmod 4$.
- If $x \equiv 3 \pmod 4$, then $f(x) = (4\tilde{x}+3)f_1 + (4\tilde{x}+3)^2 f_2 \pmod{K} \equiv 3f_1 + f_2 \pmod 4$, for some integer $\tilde{x}$. Since $f_2$ is even and $f_1$ is odd (see item 1) of Definition 1), $f(x) \pmod 4$ is odd. Also, $3f_1 + f_2 \not\equiv f_1 + f_2 \pmod 4$, and the result follows. ∎

The bound in Theorem 2 is based on the upper and lower constituent codewords in Figs. 10 and 11. For details, see the proof of Theorem 2 in Appendix A. Now, consider the codeword in Fig. 10. Let $x \in \mathbb{Z}_K$ denote the leftmost systematic 1-position in the upper constituent codeword. As shown in the proof of Theorem 2, we require $x \equiv 1 \pmod 4$. Also, for a given value of $x$ such that $x \equiv 1 \pmod 4$, the fundamental paths in Fig. 10 may wrap around at the end of trellis. Since all the systematic 1-positions are determined by the value of $x$, there will be exactly $L_i - 1$ values for $x$ that will make the $i$th, $i = 0, \ldots, Q-1$, fundamental path wrap around at the end of the trellis, where $L_i$ is the length of the $i$th fundamental path and $Q$ is the total number of fundamental paths in the upper and lower constituent codewords. By repeating the argument, we get that there will be at most $L - Q$ values for $x$ that will make at least one of the fundamental paths wrap around at the end of the trellis, where $L = \sum_{i=0}^{Q-1} L_i$. Since if $x_1 \equiv x_2 \pmod 4$, then $f(x_1) \equiv f(x_2) \pmod 4$ for any $x_1, x_2 \in \mathbb{Z}_K$ and any QPP $f(x) = f_1 x + f_2 x^2 \pmod{K}$ (see Lemma 3 above), the condition that a fundamental path should not wrap around at the end of the trellis, will remove at most one more value of $x \equiv 1 \pmod 4$ than any other value for $x$. Thus, we get the condition $K - (L - Q) \geq 3(Q + 1) + 1$, which simplifies to $K \geq L + 2Q + 4 = 112$, since, for the codeword in Fig. 10, $Q = 10$ and $L = 88$.

Note that the argument above can be repeated for the codeword in Fig. 11, and we get the numerical value of $K \geq 120$. Thus, if $K \geq \max(112, 120) = 120$, there will exist at least one value for $x$ such that $x \equiv 1 \pmod 4$ and such that none of the fundamental paths in Figs. 10 and 11 will wrap around at the end of the trellis. Furthermore, since there is a finite number of QPPs with a quadratic inverse for each value of $K$ in the range $[33, 119]$ and such that $4|K$ and the conditions in (10) are satisfied, it can be numerically checked that there will always exist, for any QPP $f(x) = f_1 x + f_2 x^2 \pmod{K}$ and its inverse $g(x) = g_1 x + g_2 x^2 \pmod{K}$, at least one value for $x$ such that $x \equiv 1 \pmod 4$ and such that none of the fundamental paths in Fig. 10 (when $f_1 + f_2 \equiv g_1 + g_2 \equiv 1 \pmod 4$) and in Fig. 11 (when $f_1 + f_2 \equiv g_1 + g_2 \equiv 3 \pmod 4$) will wrap around at the end of the trellis. Also, for $K = 32$, this is not the case, and Lemma 2 is proved.

## ACKNOWLEDGMENT

The authors would like to thank the anonymous reviewers for their helpful comments and suggestions.